\documentclass[fleqn,usenatbib]{mnras}

\usepackage{newtxtext,newtxmath}

\usepackage[T1]{fontenc}

\DeclareRobustCommand{\VAN}[3]{#2}
\let\VANthebibliography\thebibliography
\def\thebibliography{\DeclareRobustCommand{\VAN}[3]{##3}\VANthebibliography}

\usepackage{graphicx}	
\usepackage{amsmath}	

\title[Rapid magnetic field decay in RX J0720.4-3125]{Indication of rapid magnetic field decay in X-ray Dim Isolated Neutron Star RX J0720.4-3125}

\author[Andrei Igoshev \& Sergei Popov]{
Andrei P. Igoshev,$^{1}$\thanks{E-mail: a.igoshev@leeds.ac.uk}
\& Sergei B. Popov$^{2}$ \\
$^{1}$Department of Applied Mathematics, University of Leeds, LS2 9JT Leeds, UK\\
$^{2}$Abdus Salam International Center for Theoretical Physics 34151, Strada Costiera 11,  Trieste, Italy
}

\date{Accepted XXX. Received YYY; in original form ZZZ}

\pubyear{\the\year{}}

\begin{document}
\label{firstpage}
\pagerange{\pageref{firstpage}--\pageref{lastpage}}
\maketitle

\begin{abstract}
Magnetic field evolution of neutron stars is a long-standing debate.  
The rate of magnetic field decay for isolated, non-accreting neutron stars can be quantified by measuring the negative second derivative of the spin period. Alternatively, this rate can be estimated by observing an excess of thermal emission with respect to the standard cooling without additional heating mechanisms involved.
 One of the nearby cooling isolated neutron stars -- RX J0720.4-3125, -- offers a unique opportunity to probe the field decay as for this source there are independent measurements of the surface X-ray luminosity, the second spin period derivative, and magnetic field. 
We demonstrate that the evolution rate of the spin period derivative is in correspondence with the rate of dissipation of magnetic energy of the dipolar field if a significant part of the released energy is emitted in X-rays. 
The instantaneous time scale for the magnetic field decay is $\sim 10^4$~years. 
\end{abstract}




\section{Introduction} \label{sec:intro}

X-ray Dim Isolated Neutron Stars (also known as the Magnificent Seven, hereafter M7) are a group of nearby thermally emitting neutron stars (NSs) observed in X-rays and optics with no detected radio emission (for recent reviews see \citealt{BorgheseEsposito2023hxga, 2023Univ....9..273P}). These objects do not demonstrate transient behaviour. Their thermal X-ray spectra show no robust evidence of a power-law tail. These features set M7 apart from Galactic magnetars which have similar spin periods, and severely limit the possible role of the magnetosphere in producing X-ray emission and timing irregularities in the case of M7.  These sources have strong dipolar magnetic fields with typical value $\gtrsim 10^{13}$~G. Surface temperatures of M7 are higher than expected based on their ages without additional heating (e.g., \citealt{2018A&A...609A..74P}). Their surface thermal luminosities are also higher than their spin-down luminosities. Thus, it is often assumed that they are heated due to magnetic field decay and represent descendants of magnetars (e.g. \citealt{2010MNRAS.401.2675P, 2013MNRAS.434..123V}).

Recently, \cite{2024ApJ...969...53B} provided the first measurement of the second derivative of the spin period, $\ddot P=-4.1\times 10^{-25}$~s~s$^{-2}$, for one of the M7 sources -- RX J0720.4-3125 (RX0720 hereafter). This allows the authors to derive the braking index $n \equiv 2 - \ddot P P / (\dot P)^2 \approx 680$. This is much larger than the value $n=3$ expected for the magnetic dipole decelerating in the vacuum. \cite{2024ApJ...969...53B} propose that this large value is due to irregularities of the spin behaviour. Here we study an alternative explanation that this large braking index is due to a rapid magnetic field decay in this NS similar to our early analysis of large braking indices of isolated radio pulsars \citep{IgoshevPopov2020MNRAS}. 

This letter aims to show, based on independent observational evidence, that the external dipolar magnetic field of  RX0720 presently decays on a time scale $\approx 10^4$~years. Following the same logic we predict that the remaining M7 objects have braking indices around hundreds.

\section{Timescales of magnetic field decay}

In this section, at first we provide an estimate for instantaneous magnetic field decay based on the measured braking index. Second, we estimate the required magnetic field decay timescale to support thermal X-ray emission via the Ohmic heating of the crust. Then, we compare these scales.

\subsection{Decay as evidenced by the braking index}

For our purposes, the magneto-rotational evolution of NSs can be described following a simplified version of the magneto-dipole equation, see e.g., \cite{PhilippovTchekhovskoy2014MNRAS}:

\begin{equation}
    I\omega \dot \omega \propto B^2 \omega^4.
\end{equation}
Here $\omega=2\pi/P=2\pi \nu$ is the cyclic frequency, $B$ is the surface magnetic field of a NS, and $I$ -- its moment of inertia. 

It is convenient to introduce the braking index as a characteristic of spin evolution. It is defined as:
 
\begin{equation}
    n=\nu \ddot \nu /\dot \nu^2.
\end{equation}
For the spin-down with a constant magnetic field and other parameters of the NS, from eqs.~(1,2) one obtains $n=3$. A decaying magnetic field results in $n>3$. 

If the magnetic field has a non-zero first derivative but other parameters (the moment of inertia and magnetic inclination) are constant, after a simple algebra, one obtains:

\begin{equation}
    \frac{\dot \nu^2}{\nu^4}(n-3)\propto - B\dot B.
    \label{tau}
\end{equation}
Large positive braking index means negative magnetic field derivative thus translating into decay of magnetic field. In the case of RX0720 we are in the limit of $n\gg 3$, thus 
\begin{equation}
n\frac{\dot \nu^2}{\nu^4}\propto - B\dot B.    
\end{equation}

In order to get an estimate for magnetic field decay timescale, we assume that the field decays exponentially:
\begin{equation}
   B=B_0 \exp{\left(-t/\tau\right)}, 
\end{equation}
where $B_0$ is the initial field, and $\tau$ is some characteristic time scale of decay. Note, that $\tau$ by itself may depend on $B$ or other parameters, but we are interested in an instantaneous value of $\tau$.
Then we derive:
\begin{equation}
    \dot B= - \frac{B}{\tau}.
    \label{bdot}
\end{equation}

With eqs.(\ref{tau}, \ref{bdot}) we have:
\begin{equation}
    \tau = - \frac{2 \nu}{n\dot \nu} \approx 10^4 \, n_{1000} \nu_{-1} \dot \nu_{-15}^{-1}\, {\rm yrs}.
    \label{taubn}
\end{equation}
Here and below we use the convention $A_x=A/10^x.$

\subsection{Ohmic heating and X-ray radiation}

Now let us assume that all the magnetic energy released due to decay is via the Ohmic heating of the crust and is consequently emitted in X-rays to produce the luminosity $L_\mathrm{X}$. This is a simplification as some heat can be transported inwards to the core and then emitted by neutrinos (see e.g. \citealt{2014MNRAS.442.3484K}). 
For relatively low energy release, expected for M7 sources, the fraction of the energy emitted from the surface is relatively high if magnetic energy is released not very deep in the crust and direct URCA processes are not activated in the core. The existence of superfluidity in the crust helps to increase the surface temperature for the given energy release and depth \citep{2017JPhCS.932a2047K}.
In addition, there might be some luminosity contribution from the remaining heat emitted from the surface. 

The magnetic energy can be roughly estimated as:
\begin{equation}
   E_\mathrm{mag}=\left(\frac{B^2}{8\pi} \right) \left(\frac{4}{3}\pi R^3 \right) = \frac{B^2R^3}{6}. 
\end{equation}
Here $R$ is the NS radius. For our estimates, we assume $R_6=R/10^6 \mathrm{cm} = 1$. If we assume that the magnetic field is confined in the crust with volume of $V\approx 4 \pi h R^2$ and thickness $h \sim 0.3$~km then the estimate of the total magnetic energy is reduced by an order of magnitude. On the other hand, we neglect here contributions from non-dipolar (and non-poloidal) field components. Note, that for RX0720 there is evidence for a strong non-dipolar external magnetic field \citep{2017JPhCS.932a2007B}. 

Then the energy release is:
\begin{equation}
    \dot E_\mathrm{mag}=\frac{B\dot B R^3}{3}.
\end{equation}

Let us assume that $\dot E_\mathrm{mag}\equiv L_\mathrm{X}$ and similarly use eq. (\ref{bdot}): 
\begin{equation}
L_\mathrm{X} = \frac{B^2 R^3}{3\tau_\mathrm{B}}    
\label{eq:lum}
\end{equation}
which results in time scale estimate:
\begin{equation}
\tau_\mathrm{B}=\frac{B^2R^3}{3L_\mathrm{X}}\approx 10^5 B_{13}^2 L_\mathrm{X31}^{-1} {\rm yrs}.
\label{taub}
\end{equation}


It is worth noting that the magnetic field in the case of M7 objects can be estimated by two different methods. First, based on spin period and period derivative. Second, from the proton cyclotron line properties. Both estimates are in good correspondence with each other (e.g., \citealt{2007Ap&SS.308..181H}). For our estimates below we use the field values derived from the spin-down rate.

\section{The case of RX J0720.4-3125}

Measurement of the second derivative of the spin period for an NS which is supposed to be powered by magnetic field decay and for which there is an independent measurement of the magnetic field via spectral data, provides a unique opportunity to probe if its 
spin evolution and thermal surface emission are in correspondence with each other. 

RX0720 has $B_{13}=2.5$ and $\dot \nu_{-15} = -1.02$. The value of the magnetic field is derived from the magneto-dipole formula and is in good correspondence with the value $B_{13}\approx 5$ obtained from the spectral feature \citep{2004A&A...419.1077H} interpreted as caused by cyclotron resonance scattering of protons by magnetic field.
For RX0720 from eq. (\ref{taubn}) we obtain $\tau_{0720}=-(2 \nu)/(n \dot \nu)=1.1\times 10^4\, {\rm yrs}$. 
If we use this value to calculate the expected luminosity via eq.~(\ref{eq:lum}) then we obtain 
$L_\mathrm{X}\approx 6\times 10^{32}$~erg~s$^{-1}$ or $L_\mathrm{X}\approx 0.6\times 10^{32}$~erg~s$^{-1}$ if magnetic field is present in the crust only. 
Surprisingly, this is very close to the luminosity of RX0720 which is equal to $\sim (1-3)\times 10^{32}$~erg~s$^{-1}$ \citep{2020MNRAS.496.5052P}. This means that the time scale derived from the spin period evolution is in good correspondence with the scale $\tau_\mathrm{B}$ from eq.~(\ref{taub}). If we account for the fact that in this estimate we use the total volume of the NS and part of the energy can be emitted by neutrinos, then the correspondence is still good as the observed luminosity is smaller than the one derived from eq.~(\ref{eq:lum}) assuming $\tau_{0720}=\tau_\mathrm{B}$. 

\section{Discussion}

The time scale derived for RX0720 is much shorter than the kinematic age of RX0720 found to be $\sim 0.4$~-~0.5~Myr \citep{2010MNRAS.402.2369T}. Also, this scale is not in correspondence with the Hall one 
$\tau_\mathrm{Hall}\sim(10^3-10^4) B_{15}^{-1}$~yrs \citep{2008A&A...486..255A} if we substitute the external dipolar field values derived for the M7 objects. 

Numerical simulations of the magneto-thermal evolution of NSs predict that the M7 sources have large braking indices (see their nearly vertical paths in $P$~--~$\dot P$ diagram Fig.~10 by \citealt{2013MNRAS.434..123V}). Reading numerical values for $\dot P$ and ages for track with initial magnetic field $B = 3\times 10^{14}$~G from that plot we estimate the $\ddot P\approx 4\times 10^{-26}$~s s$^{-2}$ which translates to braking index of $n\approx 100$.
To explain values $n\sim 1000$ it seems to be necessary to assume an episode of a faster magnetic energy dissipation, maybe due to some instability. 

\subsection{Predictions for other M7 sources}

Recently, new timing data was obtained for another of the M7 objects --- RX J0806.4-4123 \citep{2024arXiv240704337P}. In this case, we can obtain $\tau_\mathrm{B}\approx 10^5$~yrs. We can use it to predict the values of $n$ and $\ddot \nu$ assuming $\tau=\tau_\mathrm{B}$. We obtain $n_{0806}\approx 1000$ and $\ddot \nu_{0806}\approx 6\times 10^{-29}$~Hz~s$^{-2}$. A similar procedure can be done for other M7 sources using recent data from \cite{2024ApJ...969...53B}: 
\begin{equation}
n=\frac{6\nu L_\mathrm{X}}{B^2 R^3\dot \nu}= 60 \nu_{-1} L_\mathrm{X 31} B_{13}^{-2} R_6^{-3}\dot \nu_{-15}^{-1}.
\end{equation}
Thus, we predict that for the rest of the M7 sources braking indices are about a few hundreds.
Similarly, we can predict values of $\ddot \nu$:
\begin{equation}
    \ddot \nu= -\frac{6L_\mathrm{X}\dot \nu}{B^2R^3}= 6\times 10^{-28} L_\mathrm{X31} \dot \nu_{-15} B_{13}^{-2} {\rm Hz}\, {\rm s}^{-2}.
\end{equation}
Results are presented in Table 1. 

\begin{table*}
    \centering
    \begin{tabular}{lccccc}
    \hline
    Name & $L_\mathrm{X31}$ & $\dot \nu_{-15}$ & $B_{13}$ &  $n$ & $\ddot \nu _{-28}$ \\
         & $10^{31}$~erg s$^{-1}$ & $10^{-15}$ Hz s$^{-1}$ & $10^{13}$~G & & $10^{-28}$~Hz s$^{-2}$\\
    \hline
    RX J0420-5022  & 0.6 & $-2.44$ & 1   & $40$ & $8.8$ \\
    RX J0806-4123 & 1.6 & $-0.08$ & 1.1 & 1000 & 0.6 \\
    RX J1308+2127  & 33 & $-1.05$ & 3.4   & $160$ & 16 \\
    RX J1856-3754  & 6.5 & $-0.6$ &  1.5  & $520$ & 10 \\
    RX J2143+0654  & 10 & $-0.47$ &  2 & $350$ & 7 \\
    \hline
    \end{tabular}
    \caption{Predicted braking indices and second time derivatives of frequency. The thermal X-ray luminosities $L_\mathrm{X31}$ are taken from Potekhin et al. (2020). If a range is given we take the medium value.  $B$ and $\dot \nu$ are taken from Bogdanov and Ho (2024). For RX J0806 all data are taken from Posselt et al. (2024).}
    \label{tab:predicted_n}
\end{table*}

\subsection{Excluding the fallback disk model}

\cite{Ertan2014MNRAS} proposed that the M7 objects are neutron stars surrounded by a fallback disk. In particular, these authors provide a model describing spin period, period derivative, and associated X-ray luminosities. 
The measurement of the second spin period derivative provides an opportunity to check the model by Ertan et al.

In Fig.~3 of their paper, \cite{Ertan2014MNRAS} presented plots for the evolution of spin period and period derivative. Unfortunately, their code is not publicly available. Thus, we cannot obtain the exact numbers. Instead, we extract the numerical values from their Fig.~3 and model time evolution of $\dot P$ assuming that it approximately follows a power-law at the moment when their model can reproduce the RX0720 timing properties: 
\begin{equation}
\dot P = C t^\alpha.    
\end{equation}
Here the time is measured in seconds and we estimate $\alpha = -3.8$ and $C = 8\times 10^{34}$ to be compatible with Fig.~3 of \cite{Ertan2014MNRAS}. 
For this evolution of the period derivative we obtain:
\begin{equation}
\ddot P = \alpha C t^{\alpha - 1}.    
\end{equation}
Similarly we can compute the braking index as:
\begin{equation}
n = 2 - \frac{\ddot P P}{(\dot P)^2}.    
\end{equation}
Substituting numerical values suggested by \cite{Ertan2014MNRAS} for the age of RX0720 which is taken to be equal to $1.45\times 10^5$~years (2-3 times smaller than the kinematic age mentioned above), we obtain $n\approx 80$. Although it has a correct sign, the braking index estimate is nearly an order of magnitude below the actual value presented by \cite{2024ApJ...969...53B}.


Finally, we note that the values of the magnetic field proposed for the M7 objects by \cite{Ertan2014MNRAS} are incompatible with the values estimated using the spectral features following the assumption of proton cyclotron lines.

\subsection{Possible physical mechanisms for delayed fast decay of magnetic field}

If we accept the evidence presented in the previous sections, it is worth discussing possible physical mechanisms that could cause magnetic field evolution on such an extraordinarily short time scale of $10^4$~years 
after hundreds of kyr of slower evolution.
Known estimates for the magnetic field decay due to the crust resistivity vary between 0.1-1~Myr for pasta layer in magnetars \citep{Pons2013NatPh} and $\gtrsim 10$~Myr \citep{Igoshev2019MNRAS} for normal radio pulsars with some indications of even $\approx 30$~Myr scale \citep{Igoshev2019MNRAS}. The Hall time scale for the crust-confined magnetic field is $\tau_\mathrm{Hall}\sim (10^5-10^6)$~yrs~$B_{13}^{-1}$ for magnetic field comparable to dipole estimates. 
These mechanisms for magnetic field evolution in the crust do not allow for an episode of \textit{delayed} field decay. Therefore, we must look elsewhere to identify a source of this evolution. 

The NS core has long been suggested as a potential site of delayed magnetic field evolution. For example, it is long known that many magnetic field configurations are not stable in the core, see e.g. \cite{LanderJones2012MNRAS} and references therein. It was shown that a purely poloidal magnetic field is unstable under the influence of single fluid ambipolar diffusion \citep{IgoshevHollerbach2023MNRAS}. This configuration induces new electric currents in the crust and could thus release thermal energy on time scales comparable to the ambipolar diffusion timescale. This time scale is sensitive to temperature and magnetic field strength. The ambipolar diffusion requires time for NS to cool down and instabilities to grow which might explain the late onset of the decay. The decay due to the ambipolar diffusion was already briefly discussed exactly for RX0720 by \cite{Cropper2004MNRAS}. However, these authors considered a different time scale of evolution with just a mild decay.

An alternative could be a fast evolution due to the core transition to superconductor/superfluid state \citep{Glampedakis2011MNRAS}. A very recent research by \cite{Bransgrove2024arXiv} suggests that the internal crustal magnetic fields could be amplified by orders of magnitude during the flux expulsion which should inevitably lead to enhanced 
thermal emission and accelerated evolution. Superconductor transition is sensitive to the temperature which can explain the late onset. 
However, detailed studies of the magnetic field evolution in the NS core are still in their early phase with few robust results. 
Therefore, we cannot conclusively prove that the core evolution is responsible for the enhanced X-ray luminosity of M7.

\section{Conclusions}

In the case of one of the Magnificent Seven objects RX J0720.4-3125, we demonstrate a peculiar coincidence between the time scales of magnetic field dissipation obtained from the braking index and from the X-ray luminosity. In both derivations, we assume that the field decay is the main process responsible for the anomalous spin evolution and the observed surface emission. 
We suggest that the observed thermal X-ray luminosity can be an indicator for the dipole field evolution. Thus, we make predictions for braking indices and $\ddot \nu$ of other M7 sources. Braking indices are expected to be $\sim $~a few hundred and $\ddot \nu\sim 10^{-28}-10^{-27}$~Hz~s$^{-2}$. 



\section*{Data Availability}
No new data is generated in the article beside numbers presented in Table 1.

\section*{Acknowledgements}

Authors thank anonymous referee for their comments which helped to improve the manuscript.
The work of A.I. was supported by STFC grant no.\ ST/W000873/1.
S.P. thanks the Simons Foundation for the opportunity to work at ICTP.





\bibliographystyle{mnras}
\bibliography{rxj0720} 








\bsp	
\label{lastpage}
\end{document}